\documentclass[letterpaper, 10pt, conference]{ieeeconf}

\IEEEoverridecommandlockouts
\usepackage{amsmath,amssymb,amsfonts}

\usepackage{amsthm}
\usepackage{algorithmic}
\usepackage{graphicx}
\usepackage{textcomp}
\usepackage{xcolor}
\usepackage{cite}

\newtheorem{theorem}{Theorem}
\newtheorem{lemma}{Lemma}
\newtheorem{proposition}{Proposition}

\theoremstyle{definition}

\newtheorem{assumption}{Assumption}

\theoremstyle{remark}
\newtheorem{remark}{Remark}

\title{\LARGE \bf
Composite Control of Grid-Following Inverters for
Stabilizing AI-Induced Fast Power Disturbances}

\author{Miroslav Kosanic~\IEEEmembership{Member,~IEEE},
        Marija Ili\'{c}~\IEEEmembership{Senior Member,~IEEE}
\thanks{This work has been submitted to the IEEE for possible
publication. Copyright may be transferred without notice, after
which this version may no longer be accessible.}
\thanks{This work was supported by the National Science Foundation under Grant
No. ECCS-2002570, ASCENT: Boosting Cyber and Physical Resilience of
Power Electronics-Dominated Distribution Grids. Partial support to the
second author by the MIT-GE Vernova Alliance is greatly appreciated.
The authors also thank Dr. Tong Huang for valuable discussions.
M. Kosanic is a Ph.D. candidate at the Laboratory for Information and
Decision Systems (LIDS), Massachusetts Institute of Technology, Cambridge,
MA, USA (e-mail: kosanic@mit.edu).
M. D. Ili\'{c} is a Joint EECS Adjunct Professor and a Senior Research
Scientist at the Laboratory for Information and Decision Systems (LIDS),
Massachusetts Institute of Technology, Cambridge,
MA, USA, and a Professor Emerita in the Electrical and Computer
Engineering (ECE) Department, Carnegie Mellon University, Pittsburgh,
PA, USA (e-mail: ilic@mit.edu).}}

\begin{document}

\maketitle
\thispagestyle{empty}
\pagestyle{empty}

\begin{abstract}
AI data center loads create  query-driven power transients on millisecond timescales. Such loads can violate the timescale separation assumptions underlying  internal inverter  control of grid-following resources collocated with data centers as supplementary generation. This paper develops a singular perturbation-based modeling and control  for stabilizing fast power  
imbalances. We show that physically-implementable droop control  is derived  and valid by requiring reduced-system stability rather than being imposed a priori, and that AI workloads satisfy a bounded-rate disturbance class due to physical filtering in power delivery hardware. The analysis yields explicit gain bounds linking inverter parameters to disturbance rejection performance, a modulation admissibility condition ensuring physical realizability of the feedback linearizing control, and a feasibility condition identifying the maximum tolerable load ramp rate.
Numerical simulations validate the theoretical predictions under stochastic AI transients.
\end{abstract}

% and a delay tolerance bound quantifying the tradeoff between response speed and communication latency.

\section{Introduction}
\label{sec:intro}

The rapid deployment of artificial intelligence infrastructure has 
introduced electrical loads whose power consumption differs fundamentally 
from traditional industrial loads. AI data centers exhibit query-driven GPU 
transients with ramping rates on the order of 
milliseconds~\cite{li2024unseen}, producing highly variable, nonstationary 
power demand dominated by sharp transients that conventional load models 
fail to capture.

Grid-following inverters interfacing with these loads are directly exposed 
to such fluctuations. The standard cascaded control 
architecture~\cite{geng2022dynamic} employs a fast inner current loop and a 
slower outer power regulation loop, with stability relying on sufficient 
timescale separation. Existing designs assume quasi-stationary 
disturbances~\cite{ajala2018hierarchy}, under which timescale separation holds ~\cite{guo2014dynamic}-~\cite{chang2015stability}. 
AI-driven loads violate this assumption, raising fundamental questions about 
whether classical architectures remain stabilizable and under what 
conditions performance guarantees can be maintained.

Singular perturbation methods have been applied to inverter-based systems 
for model reduction and small-signal stability 
analysis~\cite{ajala2018hierarchy,CaduffMOR2021,JafarpourSP2021,WuSPPV2024}. 
These works focus on linearized dynamics around nominal operating points with slowly varying disturbances to reduce the complexity of network dynamics. Our work differs in objective, where we employ the composite control framework of~\cite{chow1978near} to derive nonlinear, disturbance-aware stability guarantees under bounded-rate, time-varying loads representative of AI data center power profiles.

The main contributions are as follows:
\begin{enumerate}
    \item A grid-following inverter model is cast into singular perturbation 
    form, and analysis of the reduced system reveals that the droop control 
    structure emerges from stability requirements rather than being imposed 
    a priori.

    \item A bounded-rate disturbance class is introduced for loads with 
    bounded amplitude and rate of change. AI workloads are shown to belong 
    to this class due to physical filtering in power delivery hardware.

    \item Composite stability is established using singular perturbation 
    theory and ISS arguments, yielding explicit gain bounds, a modulation 
    admissibility condition ensuring physical realizability of the feedback linearizing control, and a feasibility condition identifying the maximum tolerable load ramp rate.

    % \item A delay tolerance bound quantifies the tradeoff between fast 
    % power response and robustness to communication latency.
\end{enumerate}

The remainder of this paper is organized as follows. 
Section~\ref{sec:modeling} presents the grid-following inverter model in 
the synchronous reference frame. Section~\ref{sec:SP_design} develops the 
singular perturbation analysis, derives the composite control design with 
modulation admissibility guarantees, and establishes gain bounds, 
feasibility conditions.
% , and delay constraints. 
Section~\ref{sec:AI_load} introduces the AI workload model and 
establishes that such loads satisfy the bounded-rate disturbance class. 
Section~\ref{sec:numerical} provides numerical validation under stochastic AI transients and feasibility analysis. 
Section~\ref{sec:conclusion} concludes the paper.
\section{Modeling}
\label{sec:modeling}

This section presents a model of a grid-following inverter connected to a 
stiff grid through an $RL$-filter in the synchronous $dq$ frame.

\subsection{System Description}

\begin{figure}[t]
    \centering
    \includegraphics[width=\columnwidth]{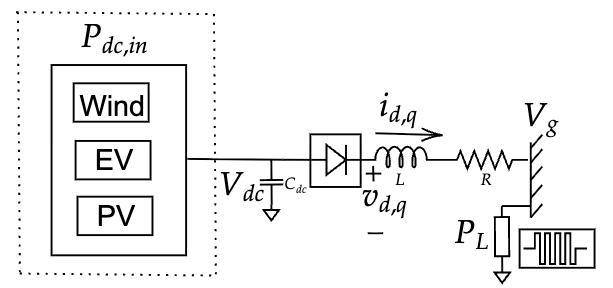}
    \caption{Grid-following inverter with local AI load connected to stiff 
    grid. The inverter regulates power through the $RL$-filter while the 
    local load $P_L(t)$ exhibits fast, query-induced transients.}
    \label{fig:system}
\end{figure}

Figure~\ref{fig:system} shows the system topology. A grid-following 
inverter interfaces a DC source with a stiff grid of voltage $V_g$ through 
an $RL$-filter with inductance $L$ and parasitic resistance $R$. A 
time-varying local load $P_L(t)$, representing an AI data center, is 
connected at the point of common coupling (PCC). The inverter regulates 
power flow while maintaining DC-link voltage $V_{dc}$.

The $dq$ reference frame is aligned with the PCC voltage so that $V_q = 0$. 
The modulation index $\mathbf{m} = (m_d, m_q)$ produces terminal voltages 
$v_d = \kappa V_{dc} m_d$ and $v_q = \kappa V_{dc} m_q$, where $\kappa > 0$ 
is a scaling factor determined by the modulation scheme. Operation in the linear modulation region requires 
$m_d^2 + m_q^2 \leq m_{\max}^2$ and beyond this limit, voltage distortion 
invalidates the linear voltage-modulation relationship assumed by the 
feedback linearizing control in Section~\ref{sec:SP_design}.

\subsection{Electrical and Power Dynamics}

The $L$-filter inductor dynamics in the synchronous $dq$ frame are
\begin{align}
    L\dot{i}_d &= -Ri_d + L\omega_g i_q + v_d - V_g
    \label{eq:id_full} \\
    L\dot{i}_q &= -Ri_q - L\omega_g i_d + v_q
    \label{eq:iq_full}
\end{align}
where the cross-coupling terms $\pm L\omega_g i_{d,q}$ arise from the 
rotating reference frame transformation.

Instantaneous power balance on the DC capacitor gives
\begin{equation}
    C_{dc} V_{dc} \dot{V}_{dc} = P_{dc,in} - \tfrac{3}{2} V_g i_d
    \label{eq:dc_energy}
\end{equation}
under unity power factor operation ($i_q = 0$). Real power at the PCC is 
measured through a first-order filter:
\begin{equation}
    \tau_p \dot{P}_m = -P_m + \tfrac{3}{2} V_g i_d - P_L(t)
    \label{eq:power_filter}
\end{equation}
where $P_L(t)$ denotes the time-varying local load disturbance. 
Equations~\eqref{eq:id_full}--\eqref{eq:power_filter} constitute the full 
plant model.
\section{Singular Perturbation and Composite Control Design}
\label{sec:SP_design}

This section exploits the multiscale structure of the plant model to design 
a composite controller that guarantees stable power regulation under 
time-varying load disturbances. The droop control structure of the slow subsystem emerges from the 
singular perturbation framework rather than being imposed a priori.

\subsection{Feedback Linearizing Control}

The electrical dynamics \eqref{eq:id_full}--\eqref{eq:iq_full} contain 
cross-coupling terms $\pm L\omega_g i_{d,q}$ that couple the $d$ and $q$ 
axes. Feedback linearization achieves exact decoupling. Define the control 
inputs as
\begin{align}
    v_d &= V_g - L\omega_g i_q + u_d
    \label{eq:vd_FBLC} \\
    v_q &= L\omega_g i_d + u_q
    \label{eq:vq_FBLC}
\end{align}
where $(u_d, u_q)$ are virtual control inputs. Substituting into 
\eqref{eq:id_full}--\eqref{eq:iq_full} yields the decoupled dynamics
\begin{align}
    L\dot{i}_d &= -Ri_d + u_d
    \label{eq:id_decoupled} \\
    L\dot{i}_q &= -Ri_q + u_q
    \label{eq:iq_decoupled}
\end{align}

This decoupling is exact provided the required terminal voltages lie within 
the inverter's linear modulation region. This condition is verified in 
Section~\ref{sec:modulation}.

\subsection{Inner Loop and Singular Perturbation Form}

The virtual inputs are chosen to impose first-order tracking where $u_d = -k_d(i_d - i_d^*)$ and $u_q = -k_q i_q$, where $k_d, k_q > 0$ are tracking gains, $i_d^*$ is a reference to be determined, and unity power 
factor operation sets $i_q^* = 0$. The closed-loop $q$-axis dynamics are 
autonomous and exponentially stable with $i_q \to 0$. The $d$-axis 
closed-loop dynamics become
\begin{equation}
    L\dot{i}_d = -(k_d + R)(i_d - i_d^*) - Ri_d^*
    \label{eq:id_CL}
\end{equation}

Defining the small parameter $\mu := L/(k_d + R)$, the $d$-axis dynamics 
take the standard singular perturbation form:
\begin{equation}
    \mu \dot{i}_d = -(i_d - i_d^*)
    \label{eq:SP_form}
\end{equation}

\subsection{Boundary Layer Stability}

Introducing the stretched time $\tau := t/\mu$ and treating $i_d^*$ as 
constant on the fast timescale yields the boundary layer dynamics
\begin{equation}
    \frac{di_d}{d\tau} = -(i_d - i_d^*)
    \label{eq:BL}
\end{equation}

\begin{lemma}
\label{lem:BL}
For any fixed $i_d^*$, the equilibrium $i_d = i_d^*$ of \eqref{eq:BL} is 
globally exponentially stable with 
$|i_d(t) - i_d^*| = e^{-(k_d+R)t/L}|i_d(0) - i_d^*|$.
\end{lemma}

\subsection{Slow Manifold Design}
\label{sec:slow_manifold}

On the slow timescale, treating $i_d$ as an input, the power measurement 
dynamics~\eqref{eq:power_filter} form the open-loop slow subsystem:
\begin{equation}
    \tau_p \dot{P}_m = -P_m + \tfrac{3}{2} V_g i_d - P_L(t)
    \label{eq:slow_OL}
\end{equation}

Consider a general slow manifold $i_d^* = \Phi(P_m, P^*)$ where $P^*$ is a 
power setpoint from a higher layer of hierarchy. Setting $\mu = 0$ enforces 
$i_d = \Phi(P_m, P^*)$. Local stability of the resulting reduced system 
requires
\begin{equation}
    \frac{\partial \Phi}{\partial P_m} < \frac{2}{3 V_g}
    \label{eq:Phi_constraint}
\end{equation}

The minimal linear structure satisfying \eqref{eq:Phi_constraint} while 
enabling setpoint regulation is static proportional feedback:
\begin{equation}
    i_d^* = K_{pp}(P^* - P_m)
    \label{eq:droop}
\end{equation}
for which $\partial \Phi / \partial P_m = -K_{pp} < 0 < 2/(3V_g)$ for any 
$K_{pp} > 0$.

\begin{remark}
\label{rem:droop}
Equation~\eqref{eq:droop} is a $P$-$I_d$ droop characteristic where the
reference current is set proportional to power error. The significance
is that this structure emerges from the stability requirement~\eqref{eq:Phi_constraint} rather than being imposed a
priori. Any stabilizing slow manifold must satisfy
$\partial\Phi/\partial P_m < 2/(3V_g)$, and proportional feedback is
the minimal linear structure meeting this constraint while enabling
setpoint regulation. The gain $K_{pp}$ parametrizes the family of
stabilizing slow manifolds.
\end{remark}

\subsection{Reduced System}

Setting $\mu = 0$ and substituting \eqref{eq:droop} into 
\eqref{eq:slow_OL} yields the reduced power dynamics:
\begin{equation}
    \dot{P}_m = -\frac{1}{\tau_{eff}} P_m
    + \frac{1}{\tau_{eff}} \cdot
      \frac{\tfrac{3}{2} V_g K_{pp}}{1 + \tfrac{3}{2} V_g K_{pp}} P^*
    - \frac{1}{\tau_p} P_L(t)
    \label{eq:reduced_final}
\end{equation}
where the effective time constant is
\begin{equation}
    \tau_{eff} := \frac{\tau_p}{1 + \tfrac{3}{2} V_g K_{pp}}
    \label{eq:tau_eff}
\end{equation}

\begin{lemma}
\label{lem:reduced}
The reduced dynamics \eqref{eq:reduced_final} are exponentially stable for 
any $K_{pp} > 0$, with eigenvalue 
$\lambda = -1/\tau_{eff} < 0$. The steady-state droop characteristic is
$\bar{i}_d = K_{pp}(P^* + P_L)/(1 + \tfrac{3}{2} V_g K_{pp})$.
\end{lemma}

\subsection{Composite Stability}

The DC-link dynamics~\eqref{eq:dc_energy} are driven by $i_d$ from the 
power subsystem. Since $i_d$ is bounded, the DC-link energy converges to a 
bounded value determined by power balance; the DC dynamics are stable in 
cascade.

\begin{theorem}
\label{thm:composite}
Suppose $K_{pp} > 0$ and the load disturbance satisfies
\begin{equation}
    |P_L(t)| \le \Delta_P, \qquad |\dot{P}_L(t)| \le \rho_P
    \label{eq:disturbance_bounds}
\end{equation}
Then there exists $\mu^* > 0$ such that for all $0 < \mu \le \mu^*$, the 
composite system is stable. After an initial transient of duration 
$O(\mu|\ln\mu|)$:
\[
|i_d(t) - K_{pp}(P^* - P_m(t))| = O(\mu)
\]
and the measured power satisfies
\begin{equation}
|P_m(t)| \le c_1 e^{-t/\tau_{eff}} |P_m(0)| + c_2 \Delta_P + c_3 \rho_P
\label{eq:ISS}
\end{equation}
for constants $c_1, c_2, c_3 > 0$. The DC-link voltage remains bounded.
\end{theorem}

\begin{proof}
See Appendix~\ref{app:tikhonov}.
\end{proof}

The load ramp rate $\rho_P$ enters linearly in~\eqref{eq:ISS} as slowly 
varying loads induce negligible transient amplification, while faster ramps 
produce proportionally larger deviations without affecting stability.

\subsection{Modulation Admissibility}
\label{sec:modulation}

The feedback linearizing control \eqref{eq:vd_FBLC}--\eqref{eq:vq_FBLC} 
requires terminal voltages that must be realizable through PWM. With 
$i_q \approx 0$ after the boundary layer transient, the modulation indices 
are
\begin{equation}
    m_d = \frac{V_g - k_d e_d}{\kappa V_{dc}}, \qquad
    m_q = \frac{L\omega_g i_d}{\kappa V_{dc}}
    \label{eq:mod_indices}
\end{equation}
where $e_d := i_d - i_d^*$ is the tracking error. Operation in the linear 
modulation region requires $m_d^2 + m_q^2 \leq m_{\max}^2$.

\begin{proposition}
\label{prop:modulation}
Suppose the conditions of Theorem~\ref{thm:composite} hold and 
$V_{dc}(t) \geq V_{dc,\min} > 0$ for all $t \geq 0$. Define the 
worst-case error $\bar{e} := \Delta i_{\max}$ and worst-case current 
$\bar{i}_d := K_{pp}(P^* + \Delta_P)/(1 + \tfrac{3}{2}V_g K_{pp}) 
+ \bar{e}$. If
\begin{equation}
    (V_g + k_d \bar{e})^2 + (L\omega_g \bar{i}_d)^2 
    \leq (\kappa\, m_{\max}\, V_{dc,\min})^2
    \label{eq:mod_condition}
\end{equation}
then the modulation indices satisfy $\|\mathbf{m}(t)\| \leq m_{\max}$ for 
all $t \geq 0$, and the feedback linearizing control remains in the linear 
modulation region along all trajectories of the composite system.
\end{proposition}

\begin{proof}
See Appendix~\ref{app:modulation}.
\end{proof}

\begin{remark}
\label{rem:selfconsistency}
Proposition~\ref{prop:modulation} closes a potential feasibility 
circularity. Theorem~\ref{thm:composite} requires exact feedback 
linearization, which requires non-saturation, which 
Proposition~\ref{prop:modulation} verifies from the trajectory bounds 
that Theorem~\ref{thm:composite} produces. The argument is 
self-consistent because the ISS bounds are \emph{a priori} estimates 
independent of the modulation constraint. The remaining hypothesis 
$V_{dc}(t) \geq V_{dc,\min}$ holds when $P_{dc,in}(t)$ is 
lower-bounded and the DC source can supply the demanded power. Establishing this rigorously via a forward-invariance argument for \eqref{eq:dc_energy} is reserved for the journal version.
\end{remark}

\subsection{Gain Bounds}

The inner loop gain $k_d$ must satisfy lower and upper bounds. For valid 
singular perturbation with separation factor $\alpha \gtrsim 10$:
\begin{equation}
    k_d \ge k_d^{SP} := \frac{\alpha L}{\tau_p}
    \label{eq:kd_SP}
\end{equation}
For bounded quasi-steady tracking error $|e_{qs}| \le e_{\max}$ under load 
ramps:
\begin{equation}
    k_d \ge k_d^{ramp} := 
    \frac{L K_{pp} \rho_P}{(1 + \tfrac{3}{2} V_g K_{pp}) e_{\max}}
    \label{eq:kd_ramp}
\end{equation}
The modulation margin $H_{\min} := \kappa m_{\max} V_{dc,\min} - \sqrt{2}\,V_g$
and switching frequency impose upper bounds:
\begin{equation}
    k_d \le k_d^{volt} := \frac{H_{\min}}{\Delta i_{\max}}, \qquad
    k_d \le k_d^{bw} := \frac{L \omega_{sw}}{n}
    \label{eq:kd_upper}
\end{equation}
with $n \in [5, 10]$. Derivations appear in Appendix~\ref{app:gains}. The combined bound is
\begin{equation}
    \max(k_d^{SP}, k_d^{ramp}) \le k_d \le \min(k_d^{volt}, k_d^{bw})
    \label{eq:kd_combined}
\end{equation}
The bound $k_d^{volt}$ is the scalar $d$-axis restriction of the vector 
modulation condition~\eqref{eq:mod_condition}, which dominates when 
$L\omega_g \bar{i}_d \ll V_g$. The slow-side timescale separation 
requirement $\tau_{eff} \ge \alpha \mu$ yields
\begin{equation}
    K_{pp} \le K_{pp}^{SP} := 
    \frac{2}{3 V_g}\left(\frac{k_d \tau_p}{\alpha L} - 1\right)
    \label{eq:Kpp_SP}
\end{equation}

\subsection{Feasibility Condition}

A feasible controller exists if and only if the gain 
intervals~\eqref{eq:kd_combined} are nonempty, requiring the 
hardware-disturbance compatibility condition:
\begin{equation}
    \rho_P \le \rho_P^{\max} := 
    \frac{(1 + \tfrac{3}{2} V_g K_{pp})\, e_{\max}\, k_d^{\max}}{L K_{pp}}
    \label{eq:rho_max}
\end{equation}
where $k_d^{\max} := \min(k_d^{volt}, k_d^{bw})$. This links control 
design to hardware limits as for a given inverter, there exists a maximum load ramp rate compatible with acceptable tracking error.

% \subsection{Communication Delay}
% \label{sec:delay}

% In practical implementations, communication latency introduces a delay $\tau_d > 0$ in the feedback path,  such that $i_d^* = K_{pp}(P^* - P_m(t - \tau_d))$. Substituting into \eqref{eq:slow_OL} yields a delay differential equation whose stability depends on the delay magnitude.

% \begin{lemma}
% \label{lem:delay}
% The delayed reduced system is asymptotically stable if
% \begin{equation}
%     \tau_d < \tau_d^{\max} := \frac{\tau_p}{\tfrac{3}{2} V_g K_{pp}}
%     \label{eq:tau_d_max}
% \end{equation}
% \end{lemma}

% \begin{proof}
% See Appendix~\ref{app:delay}.
% \end{proof}

% The combined constraint on $K_{pp}$ incorporating timescale separation and 
% delay robustness is
% $0 < K_{pp} \le \min(K_{pp}^{SP},\; 2\tau_p/(3 V_g \tau_d))$.

% \begin{remark}
% \label{rem:tradeoff}
% Increasing $K_{pp}$ accelerates power response (smaller $\tau_{eff}$) but 
% reduces delay tolerance. The product constraint 
% $K_{pp} \cdot \tau_d < 2\tau_p/(3 V_g)$ quantifies this tradeoff, as faster 
% local response requires lower-latency communication infrastructure.
% \end{remark}
\section{AI Workload Model and Deterministic Bounds}
\label{sec:AI_load}

This section establishes that AI-driven loads satisfy the bounded-rate 
disturbance class~\eqref{eq:disturbance_bounds} required by 
Theorem~\ref{thm:composite}, despite their stochastic and irregular nature.

The net power drawn at the PCC decomposes as 
$P_L(t) = P_{base}(t) + P_{AI}(t)$, where $P_{base}(t)$ is slowly varying 
facility load and $P_{AI}(t)$ captures dynamic GPU/TPU consumption. The 
AI-related power is modeled as the output of a stable linear filter driven 
by stochastic workload arrivals:
\begin{equation}
    \dot{z}(t) = A_z z(t) + B_z w(t), \qquad P_{AI}(t) = C_z z(t)
    \label{eq:AI_filter}
\end{equation}
where $A_z$ is Hurwitz and $w(t) = \sum_{k=1}^{N(t)} b_k\,\delta(t - t_k)$ is a compound Poisson process with arrival rate $\lambda$, bounded batch sizes $b_k \in [0, b_{\max}]$, and $N(t) \sim \text{Poisson}(\lambda)$. The Hurwitz filter encodes hardware-imposed power limits, with rise and fall times of $P_{AI}(t)$ that obey physical constraints from thermal dynamics and power 
delivery networks.

\begin{assumption}
\label{ass:load}
The workload satisfies $|w(t)| \leq M_w$ a.s., and the base load satisfies 
$|P_{base}(t)| \leq M_b$, $|\dot{P}_{base}(t)| \leq \rho_b$ for all 
$t \geq 0$.
\end{assumption}

\begin{lemma}
\label{lem:AI_bounds}
Suppose Assumption~\ref{ass:load} holds and $A_z$ is Hurwitz. Then 
$P_L(t)$ satisfies
\begin{equation}
    |P_L(t)| \le \Delta_P, \qquad |\dot{P}_L(t)| \le \rho_P
    \label{eq:PL_bounds}
\end{equation}
for finite constants $\Delta_P := M_b + \|C_z\|M_z$ and 
$\rho_P := \rho_b + \|C_z A_z\|M_z + \|C_z B_z\|M_w$, where 
$M_z := \sup_{t}\|z(t)\| < \infty$.
\end{lemma}

\begin{proof}
Since $A_z$ is Hurwitz and $|w(t)| \leq M_w$, boundedness of $z(t)$ 
follows from BIBO stability. The amplitude bound uses the triangle 
inequality on $P_L = P_{base} + C_z z$. The rate bound follows from 
differentiating $\dot{P}_L = \dot{P}_{base} + C_z A_z z + C_z B_z w$, 
and bounding each term.
\end{proof}

Combining Lemma~\ref{lem:AI_bounds} with Theorem~\ref{thm:composite} 
yields the main result. The grid-following inverter with composite control is stable under AI-induced load transients, with tracking errors bounded by~\eqref{eq:ISS} where $\Delta_P$ and $\rho_P$ are the physical constants from Lemma~\ref{lem:AI_bounds}. The key insight is that hardware filtering enforces deterministic bounds on magnitude and ramp 
rate despite the stochastic nature of AI workloads.
\section{Numerical Validation}
\label{sec:numerical}

The inverter is rated at 20~kW, sized to serve two to three 
H100 GPU nodes at median utilization~\cite{latif2024empirical}. 
Parameters are summarized in Table~\ref{tab:params} and were obtained 
by the following sequential design procedure. First, hardware constraints fix the upper gain bound $k_d^{volt} = H_{\min}/\Delta i_{\max} = 
130.8/100 = 1.31~\Omega$ and $k_d^{bw} = L\omega_{sw}/n = 12.6~\Omega$ 
(with $n = 10$), giving $k_d^{\max} = 1.31~\Omega$. Second, the singular 
perturbation lower bound is $k_d^{SP} = \alpha L/\tau_p = 1.0~\Omega$, 
setting $k_d = 1.2~\Omega$ satisfies $k_d^{SP} \le k_d \le k_d^{\max}$. 
Third, the timescale separation constraint~\eqref{eq:Kpp_SP} yields 
$K_{pp}^{SP} = 0.48$~mA/W, setting $K_{pp} = 0.4$~mA/W provides margin. 
Lastly, feasibility of the ramp-tracking bound~\eqref{eq:kd_ramp} is 
verified for the target ramp rate $\rho_P$ using the chosen $k_d$ and 
$K_{pp}$.

\begin{table}[t]
\centering
\caption{System and Control Parameters}
\label{tab:params}
\begin{tabular}{llll}
\hline
\textbf{Parameter} & \textbf{Value} & \textbf{Parameter} & \textbf{Value} \\
\hline
$L$ & 2 mH        & $k_d$ & 1.2 $\Omega$ \\
$R$ & 0.1 $\Omega$ & $K_{pp}$ & 0.4 mA/W \\
$V_g$ & 277 V      & $\alpha$ & 10 \\
$V_{dc}$ & 1200 V  & $e_{\max}$ & 10 A \\
$V_{dc,\min}$ & 1100 V & $\Delta i_{\max}$ & 100 A \\
$C_{dc}$ & 10 mF   & $P^*$ & 20 kW \\
$\tau_p$ & 20 ms    & $f_{sw}$ & 10 kHz \\
$\kappa$ & 0.5      & $m_{\max}$ & 0.95 \\
\hline
\end{tabular}
\end{table}

With these parameters, the fast time constant is $\mu = L/(k_d + R) = 1.54$~ms, the effective slow time constant is $\tau_{eff} = 17.2$~ms, and the timescale ratio $\tau_{eff}/\mu = 11.1$ exceeds $\alpha = 10$. The inner loop gain satisfies $k_d = 1.2~\Omega > k_d^{SP} = 1.0~\Omega$.
% The delay tolerance is $\tau_d^{\max} = 120$~ms. 
Modulation admissibility from Proposition~\ref{prop:modulation} is verified, and the post-transient worst-case modulation magnitude is $\|\mathbf{m}\|_{\max} = 0.527$, while during the boundary layer transient $\|\mathbf{m}\|_{\max} = 0.739$, both well below $m_{\max} = 0.95$. Each scenario is initialized at steady state as determined by the reduced system~\eqref{eq:reduced_final}, and the settling time constant of the inverter output is confirmed to match $\tau_{eff} = 17.2$~ms across all scenarios.

In all simulations, the power balance at the point of common coupling is $P_L = P_{inv} + P_{net}$, where $P_{inv} = \tfrac{3}{2}V_g i_d$ is the inverter output and $P_{net} = P_L - P_{inv}$ is the power drawn from the stiff grid.

\subsection{AI Load Robustness}

The AI workload is modeled as a compound Poisson process in Section~\ref{sec:AI_load}, with base load $P_{base} = 10$~kW, arrival rate $\lambda = 50$~events/s, maximum burst amplitude $b_{\max} = 10$~kW, and filter time constant $\tau = 20$~ms. The base load represents two H100 GPU nodes at median utilization, consistent with measured node-level power demand of approximately 8~kW per 8-GPU node~\cite{latif2024empirical}. The burst amplitude corresponds to the 30-60\% power swings reported during compute-communication phase transitions~\cite{li2024unseen}, and the filter time constant reflects the 20-50~ms lag observed in power supply unit response to GPU-side load steps~\cite{li2025ai}.

\begin{figure}[t]
    \centering
    \includegraphics[width=\columnwidth]{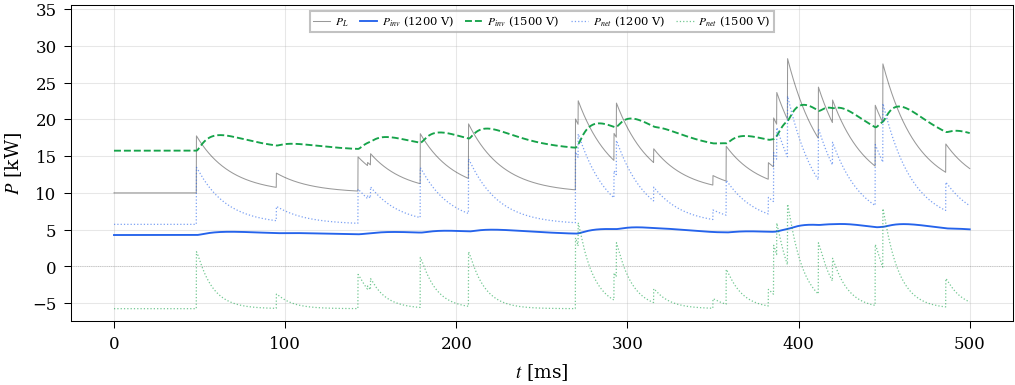}
    \caption{AI load power balance, baseline ($V_{dc}^{nom} = 1200$~V) vs.\ high-voltage design ($V_{dc}^{nom} = 1500$~V). The higher $K_{pp}$ causes the inverter to exceed the load demand ($P_{net} < 0$), exporting surplus power to the grid. Anti-windup or setpoint scheduling is required to enforce the power rating at high droop gains.}
    \label{fig:ai_comparison}
\end{figure}

Figure~\ref{fig:ai_comparison} presents the power balance under this stochastic load for both the baseline design ($V_{dc}^{nom} = 1200$~V) and the high-voltage design ($V_{dc}^{nom} = 1500$~V) introduced in Section~\ref{sec:feasibility_tradeoffs}. The load trace $P_L(t)$ exhibits rapid, irregular bursts characteristic of inference workloads, with rise times governed by the filter time constant $\tau_{rise} = 5$~ms so that the peak ramp rate $b_{\max}/\tau_{rise} = 2$~MW/s remains within the theoretical bound $\rho_P^{\max}$. With the baseline design, the inverter output $P_{inv}$ tracks the load variations within the bandwidth set by $\tau_{eff}$, with a mean contribution of 4.9~kW (approximately 34\% of the mean load), while the grid absorbs the remaining demand with a mean of $P_{net} = 9.5$~kW. Despite the stochastic fluctuations, current tracking is maintained throughout, validating Lemma~\ref{lem:AI_bounds} and the composite stability guarantee.

The limited load share is structural. At equilibrium, the reduced system~\eqref{eq:reduced_final} yields the steady-state inverter power
\begin{equation}
    \label{eq:Pinv_ss}
    P_{inv}^{ss} = \frac{1.5\,V_g\,K_{pp}}{1 + 1.5\,V_g\,K_{pp}}\,(P^* + P_L),
\end{equation}
so that the marginal participation $\partial P_{inv}^{ss}/\partial P_L = 1.5\,V_g\,K_{pp}/(1 + 1.5\,V_g\,K_{pp})$ equals 0.14 with the baseline parameters. The low droop gain $K_{pp} = 0.4$~mA/W, constrained by the modulation margin $H_{\min} = 130.8$~V analyzed in Section~\ref{sec:feasibility_tradeoffs}, limits the inverter to picking up only 0.14~kW per 1~kW increase in load while the grid absorbs the remainder. Achieving $P_{inv}$ to fully support $P_L$ would require integral action on the droop law, setpoint scheduling with $P^*$ tracking forecasted $P_L$, or grid-forming control, each of which introduces additional dynamics beyond the two-timescale framework analyzed here.

\subsection{Feasibility Region and Design Tradeoffs}
\label{sec:feasibility_tradeoffs}

Figure~\ref{fig:feasibility} visualizes the feasibility region for $k_d$ as a function of load ramp rate $\rho_P$ for both the baseline design ($V_{dc}^{nom} = 1200$~V) and a high-voltage design ($V_{dc}^{nom} = 1500$~V). For each design, the lower bounds $k_d^{SP}$ (timescale separation) and $k_d^{ramp}$ (ramp tracking) define the minimum required gain, while $k_d^{volt}$ (voltage saturation) defines the maximum.

For the baseline design, the singular perturbation bound dominates for $\rho_P < \rho_P^{crit} = 14.6$~MW/s; the ramp-rate bound becomes active above this threshold and closes the feasible region at approximately $\rho_P = 19$~MW/s. The binding constraint is the voltage saturation bound $k_d^{volt}$, where voltage margin available for feedback control is $H_{\min} = \kappa m_{\max} V_{dc,\min} - \sqrt{2}\,V_g = 130.8$~V, representing only 25\% of the total modulation capacity $\kappa m_{\max} V_{dc,\min} = 522.5$~V. The remaining 75\% is consumed matching the grid voltage.

Increasing the DC-link voltage to $V_{dc}^{nom} = 1500$~V expands the feasible design space substantially. The design procedure of Section~\ref{sec:SP_design} yields $V_{dc,\min} = 1399$~V, $H_{\min} = 273$~V, $k_d = 2.30~\Omega$, $K_{pp} = 2.7$~mA/W, and $\tau_{eff} = 9.5$~ms. As Fig.~\ref{fig:feasibility} shows, the higher $k_d^{volt}$ raises the ceiling of the feasible region. However, the larger $K_{pp}$ steepens the ramp-rate bound, reducing the maximum tolerable ramp rate from approximately 19~MW/s to 10.8~MW/s.
% and tightening the delay bound from $\tau_d^{\max} = 120$~ms to $\tau_d^{\max} = 18.1$~ms.
This reveals a fundamental tradeoff intrinsic to the gain structure, as higher DC-link voltage enables greater power response capability at the cost of reduced robustness to fast load transients.
% and tighter delay tolerance.

\begin{figure}[t]
    \centering
    \includegraphics[width=\columnwidth, height=3cm]{figures/fig_feasibility.png}
    \caption{Feasibility region for $k_d$ vs.\ load ramp rate $\rho_P$, overlaid for the baseline ($V_{dc}^{nom} = 1200$~V, blue) and high-voltage ($V_{dc}^{nom} = 1500$~V, green) designs. Higher DC-link voltage raises $k_d^{volt}$ but steepens $k_d^{ramp}$, closing the feasible region at a lower ramp rate.}
    \label{fig:feasibility}
\end{figure}
The high-voltage design's impact on power sharing is visible in Fig.~\ref{fig:ai_comparison}. With the larger $K_{pp}$, the inverter is driven well beyond load-matching, the mean inverter output of 18.0~kW exceeds the mean load, resulting in reverse power flow ($P_{net} < 0$) where the inverter exports surplus power to the stiff grid. This over-response is a direct consequence of the unconstrained droop law $i_d^* = K_{pp}(P^* - P_m)$, which commands current injection proportional to the measurement error regardless of the actual load level. At the 10~kW base load equilibrium, the commanded current reaches $i_d = 37.9$~A, corresponding to $P_{inv} = 15.7$~kW, well above both the load and the 20~kW inverter rating. In practice, this requires an explicit current saturation $i_d^* \leq P_{rated}/(1.5\,V_g) = 48.1$~A to enforce the power rating, combined with anti-windup logic on the droop law when saturation is active. Alternatively, the droop setpoint $P^*$ can be scheduled to match the expected load level, avoiding the large measurement error that drives the over-response. The simulation without saturation is presented here to expose the structural limitation of proportional droop at high gain, which the feasibility analysis does not capture because it assumes operation near the design equilibrium.
\section{Conclusion}
\label{sec:conclusion}
This paper developed a singular perturbation framework for grid-following 
inverter control under AI-driven load transients. The composite control 
structure provides rigorous stability guarantees through Theorem~\ref{thm:composite}, where the droop architecture emerges from reduced-system stability requirements. Proposition~\ref{prop:modulation} ensures physical realizability by verifying that bounded trajectories remain within the linear modulation region. Lemma~\ref{lem:AI_bounds} establishes that AI workloads satisfy the required bounded-rate disturbance class despite their stochastic nature, and the feasibility condition~\eqref{eq:rho_max} identifies the maximum tolerable load ramp rate for a given inverter design, with the 
modulation margin $H_{min}$ emerging as the binding hardware 
constraint on achievable control bandwidth and power response. Extension 
to multi-unit coordination via interaction variable framework and 
relaxation of the stiff grid assumption will be our next steps.

\bibliographystyle{IEEEtran}
\bibliography{cite}

\appendix

\section{Proof of Theorem~\ref{thm:composite}}
\label{app:tikhonov}

The proof proceeds by verifying the conditions of Tikhonov's theorem for 
the closed-loop system, which takes the standard singularly perturbed form 
$\mu \dot{x}_f = g(x_f, x_s)$, $\dot{x}_s = f(x_f, x_s, w(t))$, with 
$x_f = i_d$, $x_s = [P_m, V_{dc}]^\top$, $w(t) = P_L(t)$, and
\begin{align}
    g(x_f, x_s) &= -i_d + K_{pp}(P^* - P_m) \\
    f(x_f, x_s, w) &= \begin{bmatrix}
        \frac{1}{\tau_p}\bigl(-P_m + \tfrac{3}{2}V_g i_d - P_L\bigr) \\[4pt]
        \frac{1}{C_{dc}V_{dc}}\bigl(P_{dc,in} - \tfrac{3}{2}V_g i_d\bigr)
    \end{bmatrix}
\end{align}

Setting $g = 0$ yields the unique isolated root 
$i_d = h(x_s) := K_{pp}(P^* - P_m)$, which exists for any $P_m$. The 
boundary layer system obtained by freezing $x_s$ has Jacobian 
$\partial g / \partial x_f = -1 < 0$ uniformly, so this root is a 
uniformly exponentially stable equilibrium with unit decay rate in 
stretched time. Substituting $i_d = h(x_s)$ into the slow dynamics 
produces the reduced system~\eqref{eq:reduced_final}, whose eigenvalue 
$\lambda = -1/\tau_{eff} < 0$ for any $K_{pp} > 0$ confirms exponential 
stability. Moreover, since both $i_d$ and $P_{dc,in}$ are bounded, the 
DC-link energy equation $\dot{E}_{dc} = P_{dc,in} - \tfrac{3}{2}V_g i_d$ 
is driven by bounded inputs, ensuring that $V_{dc}$ remains bounded as 
well. Finally, $g$, $f$, and $h$ are all linear in the states, and the 
disturbance class~\eqref{eq:disturbance_bounds} guarantees 
$|P_L(t)| \leq \Delta_P$, satisfying the required smoothness and 
boundedness conditions. With all hypotheses verified, Tikhonov's theorem 
yields the existence of $\mu^*$ and the bound~\eqref{eq:ISS}. \qed

\section{Proof of Proposition~\ref{prop:modulation}}
\label{app:modulation}

The squared modulation magnitude at any time $t$ is
\[
\|\mathbf{m}(t)\|^2 
= \frac{(V_g - k_d\,e_d(t))^2 + (L\omega_g\,i_d(t))^2}
       {(\kappa\,V_{dc}(t))^2}
\]

From Theorem~\ref{thm:composite}, the tracking error satisfies 
$|e_d(t)| \leq \Delta i_{\max} =: \bar{e}$ for all $t \geq 0$, including 
the boundary layer transient where the error can reach its initial 
value. The reference current satisfies 
$|i_d^*(t)| = |K_{pp}(P^* - P_m(t))| \leq K_{pp}(P^* + \Delta_P) / 
(1 + \tfrac{3}{2}V_g K_{pp})$ from the ISS bound on $P_m$ and the 
steady-state droop characteristic of Lemma~\ref{lem:reduced}. Therefore 
$|i_d(t)| \leq |i_d^*(t)| + |e_d(t)| \leq \bar{i}_d$.

Since $V_{dc}(t) \geq V_{dc,\min}$ from the DC subsystem hypothesis, the 
numerator is maximized at its worst-case values while the denominator is 
minimized. Using $|V_g - k_d e_d| \leq V_g + k_d\bar{e}$ gives
\[
\|\mathbf{m}(t)\|^2 
\leq \frac{(V_g + k_d\bar{e})^2 + (L\omega_g\bar{i}_d)^2}
         {(\kappa\,V_{dc,\min})^2} 
\leq m_{\max}^2
\]
where the last inequality holds by hypothesis~\eqref{eq:mod_condition}. \qed

% \section{Proof of Lemma~\ref{lem:delay}}
% \label{app:delay}

% Setting $P^* = P_L = 0$ in the delayed reduced system and defining 
% $a := 1/\tau_p$ and $b := \tfrac{3}{2}V_g K_{pp}/\tau_p$, the 
% characteristic equation becomes $s + a + b\,e^{-s\tau_d} = 0$. 
% Substituting $s = j\omega$ at the stability boundary and separating real 
% and imaginary parts gives $a + b\cos\omega\tau_d = 0$ and 
% $\omega - b\sin\omega\tau_d = 0$. The first equation requires 
% $\cos\omega\tau_d = -a/b$, which admits a real solution only when 
% $b \geq a$, i.e., $\tfrac{3}{2}V_g K_{pp} \geq 1$; when this condition 
% fails, the critical delay diverges and the system is unconditionally 
% stable. Applying the Pythagorean identity 
% $\sin^2\omega\tau_d + \cos^2\omega\tau_d = 1$ to both equations yields 
% the critical frequency $\omega_c = \sqrt{b^2 - a^2}$ and the exact 
% critical delay
% \begin{equation}
%     \tau_d^{crit} = \frac{\arccos(-a/b)}{\sqrt{b^2 - a^2}}
%     \label{eq:tau_crit}
% \end{equation}

% Over the typical operating range $\tfrac{3}{2}V_g K_{pp} \in [1, 10]$, a 
% first-order expansion of~\eqref{eq:tau_crit} gives the conservative but 
% practically useful design bound 
% $\tau_d^{\max} \approx \tau_p/(\tfrac{3}{2}V_g K_{pp})$, 
% recovering~\eqref{eq:tau_d_max}. \qed

\section{Derivation of Gain Bounds}
\label{app:gains}

The bounds~\eqref{eq:kd_ramp}, \eqref{eq:Kpp_SP}, and~\eqref{eq:rho_max} 
follow from imposing $|e_{qs}| \leq e_{\max}$, $\tau_{eff} \geq \alpha\mu$, 
and nonemptiness of~\eqref{eq:kd_combined} respectively, with 
straightforward algebraic rearrangement. \qed

\end{document}